\documentclass{aa}
\usepackage{graphics}

\def\la{\lower.5ex\hbox{$\; \buildrel < \over \sim \;$}}
\def\ga{\lower.5ex\hbox{$\; \buildrel > \over \sim \;$}}

\begin{document}      
   
   \title{Ram pressure stripping of the multiphase ISM in the Virgo cluster spiral galaxy NGC~4438}

   \author{B.~Vollmer\inst{1}, M.~Soida\inst{2}, A.~Chung\inst{3}, L.~Chemin\inst{4}, J.~Braine\inst{5}, A.~Boselli\inst{6}, R.~Beck\inst{7}}

   \offprints{B.~Vollmer: bvollmer@astro.u-strasbg.fr}

   \institute{CDS, Observatoire astronomique de Strasbourg, 11, rue de l'universit\'e,
	      67000 Strasbourg, France \and
	      Astronomical Observatory, Jagiellonian University, ul. Orla 171, 30-244 Krak\'ow, Poland \and
	      Jansky fellow at NRAO, 1003 Lopezville Rd, Socorro, NM 87801, USA \and
	      Observatoire de Paris, Section Meudon, GEPI, CNRS-UMR 8111, and Universit\'e Paris 7, 
	      5 Place Janssen, 92195, Meudon, France \and
	      Laboratoire d'Astrophysique de Bordeaux (OASU), Universit\'e de Bordeaux, UMR 5804, 
	      CNRS/INSU, B.P. 89, F-33270 Floirac, France \and
	      Laboratoire d'Astrophysique de Marseille, F-13376 Marseille, France \and
	      Max-Planck-Insitut f\"{u}r Radioastronomie, Auf dem H\"{u}gel 69, 
	      53121 Bonn, Germany
              }

   \date{Received / Accepted}

   \authorrunning{Vollmer et al.}
   \titlerunning{Ram pressure stripping of the multiphase ISM in NGC~4438}

\abstract{ Ram pressure stripping of the multiphase ISM is studied in the perturbed Virgo cluster
spiral galaxy NGC~4438.
This galaxy underwent a tidal interaction $\sim 100$~Myr ago and is now strongly
affected by ram pressure stripping.
Deep VLA radio continuum observations at 6 and 20~cm are presented.
We detect prominent extraplanar emission to the west of the galactic center, which 
extends twice as far as the other tracers of extraplanar material. The spectral index of the
extraplanar emission does not steepen with increasing distance from the galaxy. This implies
in situ re-acceleration of relativistic electrons. 
The comparison with multiwavelength observations
shows that the magnetic field and the warm ionized interstellar medium traced by H$\alpha$ emission
are closely linked. The kinematics of the northern extraplanar H$\alpha$ emission,
which is ascribed to star formation, follow those of the extraplanar CO emission.
In the western and southern extraplanar regions, the H$\alpha$ measured
velocities are greater than those of the CO lines. We suggest that the ionized gas of this
region is excited by ram pressure.
The spatial and velocity offsets are consistent with a scenario where the diffuse ionized gas
is more efficiently pushed by ram pressure stripping than the neutral gas. 
We suggest that the recently found radio-deficient regions compared to $24$~$\mu$m emission
are due to this difference in stripping efficiency. 
}

\maketitle

\section{Introduction \label{sec:introduction}}

The strongly H{\sc i} deficient spiral galaxy NGC~4438 is a unique case in the Virgo cluster. 
The tidal arms of this highly
inclined galaxy are evidence of its tidal interaction $\sim 100$~Myr ago with NGC~4435 (Combes et al. 1988, 
Vollmer et al. 2005) or M~86 (Kenney et al. 2008). 
The disk of interstellar matter is severely truncated and extraplanar gas
is found to the west of the galaxy center. This extraplanar gas consists of all ISM phases:
(i) molecular hydrogen observed in CO (Combes et al. 1988, Vollmer et al. 2005), (ii) 
atomic hydrogen (Cayatte et al. 1990, Hota et al. 2007), (iii) warm ionized gas observed
in H$\alpha$ (Kenney et al. 1995, Chemin et al. 2005), (iv) hot ionized gas observed in 
X-rays (Kotanyi et al. 1983, Machacek et al. 2004), and (v) magnetic fields and
cosmic ray electrons observed in
the radio continuum (Kotanyi \& Ekers 1983, Kotanyi et al. 1983, Hota et al. 2007). 
While it is agreed that the distortion of the stellar content of NGC~4438 is due to
a tidal encounter, different mechanisms were put forward to explain the displacement of 
all ISM phases: (i) the tidal interaction which extracted the molecular gas from the center
and left it to the west of the galaxy (Combes et al. 1988), (ii) a collision between the
ISM of the two galaxies (Kenney et al. 1995), and (iii) ram pressure stripping due to the rapid
motion of NGC~4438 through the hot intracluster medium (ICM; Kotanyi et al. 1983, Chincarini
\& de Souza 1985). 

Vollmer et al. (2005) combined new CO(1--0) observations with detailed numerical simulations.
This lead to the following interaction scenario for the NGC~4438/NGC~4435 system:
NGC~4435 passed through the disk of NGC~4438  $\sim 100$~Myr ago at a radial distance of
$\sim 5-10$~kpc. The encounter was rapid ($\Delta v \sim 830$~km\,s$^{-1}$) and retrograde
(see also Combes at al. 1988). With an impact parameter $< 10$~kpc an ISM-ISM
collision is unavoidable. Its importance depends on the initial gas distributions
in NGC~4435 and NGC~4438. The current extent of NGC~4435's observed molecular gas disk is $\la 1$~kpc.
In their simulations the gas disk of NGC~4438 had an initial extent of $\sim 10$~kpc. Even 
with this initial extent the influence of an ISM-ISM collision on the final
gas distribution and velocities is low compared to that of ram pressure stripping. 
NGC~4438 evolves on an eccentric orbit within the Virgo cluster. We observe the galaxy
$\sim 10$~Myr after its closest passage to the cluster center ($\ga 300$~kpc from M87).
Ram pressure plays a key role for the evolution of the gaseous component of NGC~4438 together
with the tidal interaction. The displacement of the line profiles to higher velocities
in the southwestern region of the galaxy, the lack of CO emission in the eastern
optical disk, and the presence of double line profiles in the southwest of the galaxy center
are clear signs of ram pressure stripping. 
In the presence of a tidal interaction ram pressure stripping is more efficient, because
tidal effects have moved a significant fraction of NGC~4438's ISM to larger galactic
radii where the galactic gravitational potential is weaker (see also Vollmer 2003 and
Kapferer et al. 2008).

Recent deep H$\alpha$ observations of the M~86 region (Kenney et al. 2008) revealed a highly complex 
and disturbed ISM/ICM. NGC~4438 is connected to M~86 by several faint H$\alpha$ filaments. 
This suggests a tidal interaction between the two galaxies. Although the tidal interaction might
have occured with M~86 $\sim 100$~Myr ago, strong ongoing ram pressure is still required to explain
the CO(1--0) distribution and kinematics. 
The timescale and the relative galaxy velocities of the two scenarios (encounter with M~86 or NGC~4435)
are about the same. The higher mass of M~86 compared to that of NGC~4435 allows a larger
impact parameter for an NGC~4438 -- M~86 encounter. Since M~86 has its own intracluster medium,
the need for strong ongoing ram pressure affecting the ISM of the inner galactic disk, implies that 
ram pressure stripping due to M~86
certainly did not remove and might not have significantly affected the ISM of NGC~4438 in this region.

In this article we provide further evidence of ongoing strong ram pressure. In particular, we show that 
the warm ionized interstellar medium is stripped more efficiently than the cold dense atomic
and molecular interstellar medium.

\section{Observations\label{sec:observations}}

NGC~4438 was observed at 4850~MHz during 8~h in January 2 2006
with the Very Large Array (VLA) of the National Radio Astronomy Observatory
(NRAO)\footnote{NRAO is a facility of National Science Foundation
operated under cooperative agreement by Associated Universities, Inc.}
in the D array configuration. The band passes were $2\times 50$~MHz.
We used 3C286 as the flux calibrator and 1254+116 as the phase calibrator, the latter of
which was observed every 40~min. 
The 6~cm emission map was made using the Astronomical Imaging
Processing System (AIPS) task IMAGR with ROBUST=3.
The final cleaned map was convolved to a beam size of $18'' \times 18''$.
The bright radio source M87 caused sidelobe effects enhancing the rms noise level of NGC~4438.
We ended up with an rms level of $90$~$\mu$Jy/beam.

The 21~cm line data were obtained on January~4, 2003
with the VLA in CS configuration as part of the VIVA
survey (Chung et al. 2008). The total duration of observations was 4.8~hrs
and the pointing was centered on between NGC~4402 and
IC~3355 ($\alpha_{2000}=$12h26m29.4s, 
$\delta_{2000}=13^\circ 8^\prime 40^{\prime\prime}$),
20.7$^\prime$ away from the optical center of NGC~4438.
1331+305 (3C~286) was observed every 30 minutes as a
phase as well as a flux calibrator. The digital correlator of 
3.125~MHz bandwidth was configured to produce 127
channels and two polarizations. Online Hanning
smoothing was applied and every other channel was
discarded, yielding 63 independent channels with a 
velocity resolution of roughly 10~km\,s$^{-1}$.

The H{\sc i} data were reduced using AIPS. After flux and phase calibration,
the continuum was subtracted by making a linear fit
to the $uv$ data for a range of line-free channels at
both sides of the band using {\tt UVLIN}. The $uv$
data were imaged and cleaned using {\tt IMAGR} with 
{\tt ROBUST=1} to maximize the sensitivity while
keeping a fair spatial resolution (Briggs 1995). 
With our pointing center (between NGC~4402 and IC~3355),
NGC~4438 is found just outside of the VLA primary beam at
this wavelength, and the
image cube was corrected using {\tt PBCOR} for 
falling sensitivity at the beam edge. The total 
H{\sc i} map was produced by taking the 0th moment along
the frequency axis. The AIPS task {\tt MOMNT} creates
a mask to blank the images at a given cutoff level
and estimates moments on the full resolution blanked
cube. In creating a mask, we applied Gaussian and
Hanning smoothing spatially and in velocity,
respectively, to maximize the signal-to-noise. The total {\sc Hi}
flux that we measure is 1.94$\pm$0.24~Jy~km~s$^{-1}$,
77\% of what Cayatte et al. (1990) reported from the
observations centered on NGC~4438.

The 20~cm continuum map was created by averaging
the line-free channels. In order to reduce the
effects of interfering sources, which cause 
substantial sidelobes, we have used the AIPS
procedure {\tt PEELR}. It iteratively attempts to
calibrate on multiple fields around bright continuum
sources (self-calibration), subtract those fields
from the self-calibrated data, undo the
field-specific calibration from the residual data,
and it finally restores all fields to the residual data.
The same weighting scheme as the H{\sc i} map 
({\tt ROBUST=1}) was used for the continuum map.
The 20~cm continuum map was
flux calibrated using the observations of Condon
(1987), i.e. the VIVA data were convolved to the
resolution of Condon's map 
($54{^\prime}{^\prime}\times 54{^\prime}{^\prime}$)
and the ratio of 3.9 between Condon's and the VIVA peak
fluxes of the galaxy's central source was
applied to the VIVA 20~cm map. The final cleaned map
has a resolution of 
$17{^\prime}{^\prime}\times 15{^\prime}{^\prime}$
and an rms level of 0.2~mJy per beam.

\section{Extraplanar non-thermal radio continuum emission \label{sec:results}}

\begin{figure}
\begin{center}
        \resizebox{\hsize}{!}{\includegraphics{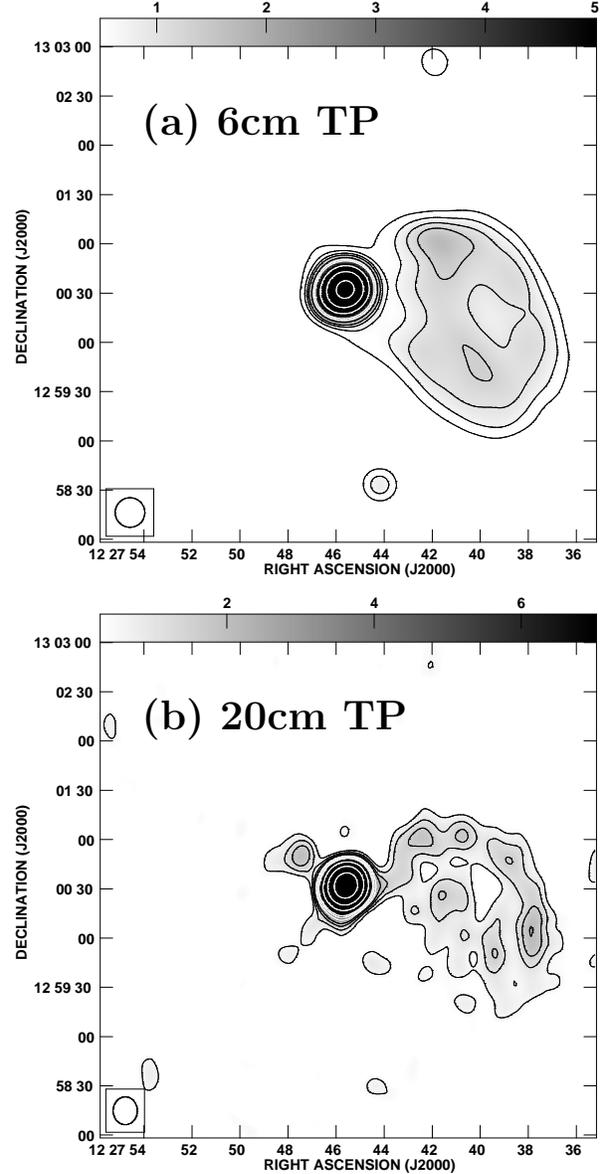}}
        \caption{Radio continuum observations of NGC~4438. (a) total power emission at 6~cm,
	  contour levels are $(1,2,3,4,6,8,10,20,40,80) \times  0.32$~mJy/beam;
	  (b) total power emission at 20~cm from H{\sc i} (VIVA), 
	  contour levels are $(1,2,3,4,6,8,10,20,40,80) \times 0.52$~mJy/beam.
	  The beam sizes are shown in the lower left corners of the images.
        } \label{fig:radio}
\end{center}
\end{figure}

At 20~cm we confirm the findings of Kotanyi \& van Gorkom (1983), i.e., we detect the galaxy center
and the western extraplanar emission with about the same surface brightness 
(0.5~mJy/beam; Fig.~\ref{fig:radio}b) as measured by these authors.
However, the morphology of our map is different from that of Kotanyi \& van Gorkom (1983).
It shows an oval structure with a position angle of about $30^{\circ}$.
We detect the same structure at 6~cm with a comparable surface brightness
(Fig.~\ref{fig:radio}a). The surface brightness within the extraplanar structure only
varies by a factor of 2, i.e. it is remarkably uniform. The hole in the center 
of the extraplanar emission might be due to emission coming from a shell rather than from a sphere.

We determine the following flux densities for the extraplanar region at 6 and 20~cm: $S_{\rm 6cm}=36$~mJy and
$S_{\rm 20cm}=32$~mJy. The flux density from Condon's 20cm image is consistent with
our measurement. At 610~MHz Hota et al. (2007) found a flux density of $S_{\rm 50cm}=160$~mJy.
This leads to a spectral index between 20 and 6~cm of $SI(20/6)=0.1$, between 50 and 20~cm
of $SI(50/20)=-1.9$, and between 50 and 6~cm of $SI(50/6)=-0.7$. 
The unusual shape of the spectral energy distribution, which varies too rapidly, 
and the high degree of polarization at 6~cm ($\sim 30$\,\%; Vollmer et al. in prep.) points to a 
substantial missing flux density at 20~cm from large scales.
This is due to missing zero-spacing information and the presence of the strong radio source M~87
close to the pointing center.
If we enhance the 20~cm flux density by a factor of $2.8$ the spectral indices between
the three frequencies all become $-0.7$. We thus conclude that the radio continuum emission
at all observed frequencies is probably non-thermal, i.e. synchrotron emission.
However, the uncertainties in the flux densities 
do not allow us to definitely exclude a thermal origin of the 6~cm emission.

The spectral index in the extraplanar regions typically steepens due to the rapid energy loss
of relativistic electrons with high energies (aging). This is observed in NGC~4522
(Vollmer et al. 2004) and the western radio lobe of NGC~4569 (Chyzy et al. 2006).
A spectral index of $-0.7$, as observed in the eastern radio lobe of NGC~4569, 
means that the relativistic electrons are re-accelerated
within the extraplanar region of NGC4438, most probably in shocks induced by ram pressure
at the surface of the shell. This scenario is consistent with the enhanced polarized emission
at 6~cm at the outer border of the extraplanar region.

\section{Multiphase ISM ram pressure stripping \label{sec:ismstripping}}

\begin{figure*}
        \resizebox{\hsize}{!}{\includegraphics{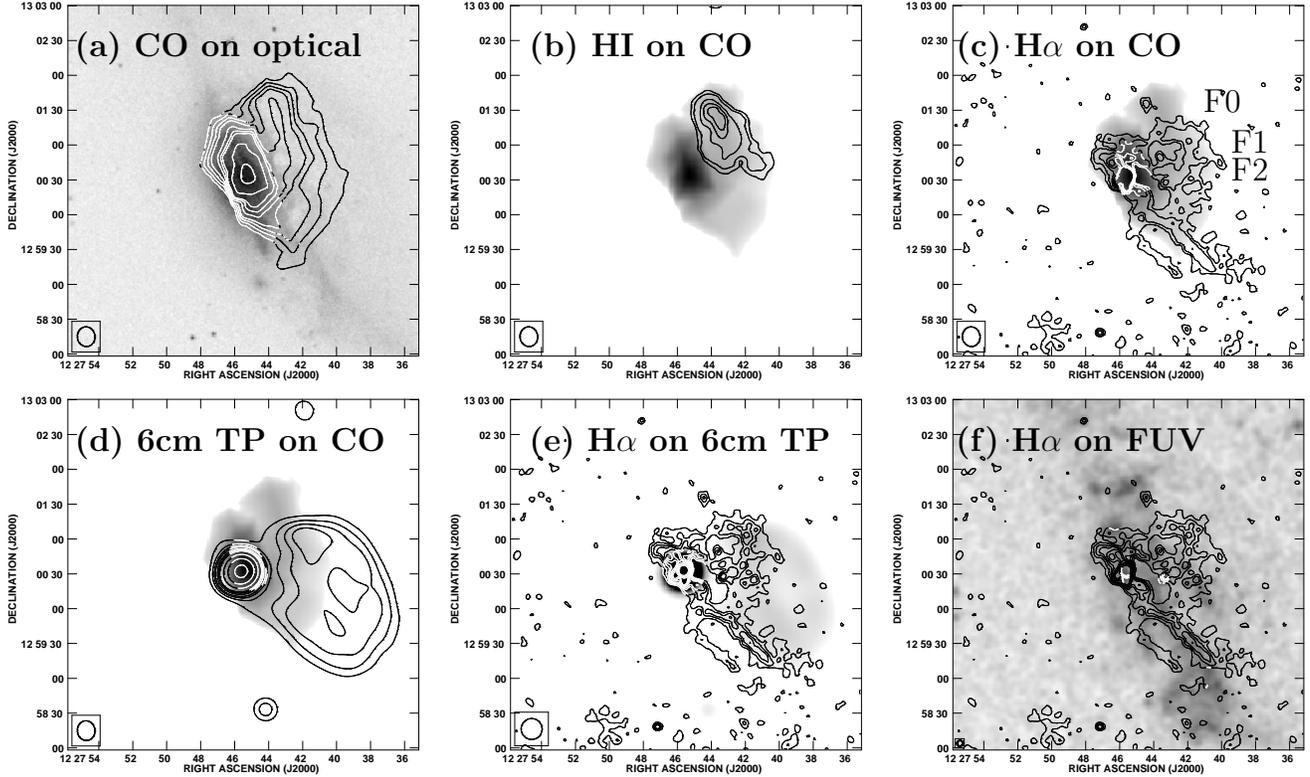}}
        \caption{
	  The different ISM phases of NGC~4438. Optical B band from DSS, CO(1-0) from Vollmer et al. (2005),
	  H{\sc i} from VIVA, H$\alpha$ from Goldmine (Gavazzi et al. 2003), 6~cm radio continuum emission
	  from this paper, GALEX FUV from Boselli et al. (2005). In panel c) the H$\alpha$ filaments 0, 1, and 2
	  mentioned in the text are labelled as F0, F1, and F2 (see also Fig.~\ref{fig:halpha}).
        } \label{fig:overlays}
\end{figure*}

\begin{figure}
        \resizebox{\hsize}{!}{\includegraphics{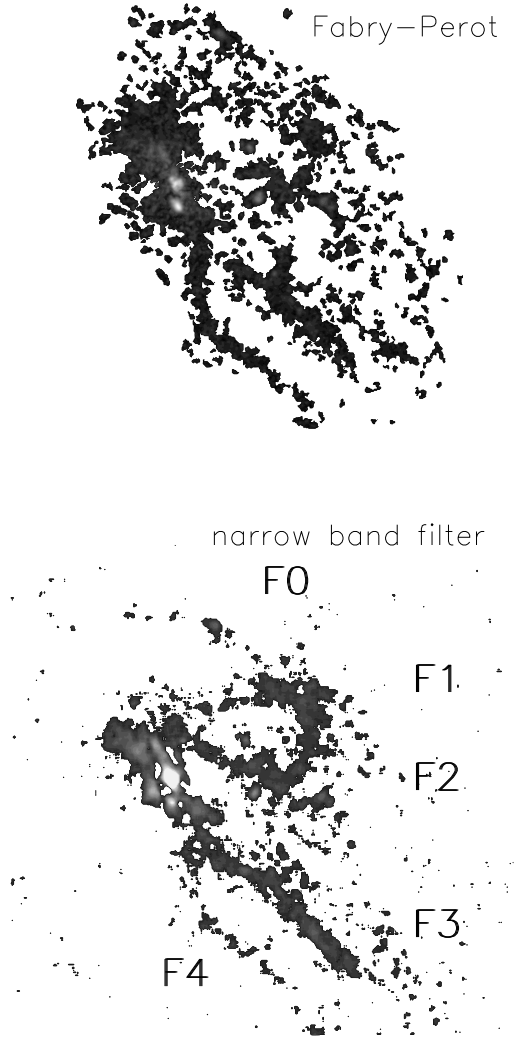}}
        \caption{Upper panel: H$\alpha$ Fabry-Perot image (Chemin et al. 2005). Lower panel:
	  H$\alpha$+[N{\sc ii}] narrow band image from Goldmine (Gavazzi et al. 2003) where a median filter of
	  7 pixels has been applied. The filaments F0--F4
	  mentioned in the text are labelled as F0--F4. Note that the greyscales are inverted
	  (bright colors correspond to high surface brightness) to enhance the visibility of
	  low surface brightness emission.
        } \label{fig:halpha}
\end{figure}

In this section we compare the spatial distribution of the different ISM phases in the
extraplanar region (Fig.~\ref{fig:overlays}). We use the following data: (i) cold molecular phase -- CO,
(ii) cool and warm atomic phase -- H{\sc i}, (iii) warm ionized phase -- H$\alpha$,
(iv) hot ionized phase -- X-rays  (Machacek et al. 2004), and (v) magnetic fields and cosmic ray electrons --
6~cm radio continuum emission.
For the H$\alpha$ emission we compare a narrow band image containing H$\alpha$ and [N{\sc ii}] emission
(from Goldmine; Gavazzi et al. 2003) to the Fabry-Perot image of Chemin et al. (2005)
which only contains the H$\alpha$ line (Fig.~\ref{fig:halpha}).
Kenney et al. (1995) describe the H$\alpha$ filaments as extended and more diffuse than the few high-brightness
H$\alpha$ knots located $\sim 1'$ west of the galaxy's nucleus (F2). The filaments' apparent widths range
between $5''$ and $15''$ ($\sim 360$~pc to 1~kpc). The emission west of the nucleus
consists of two filaments F1 and F2 (Fig.~\ref{fig:overlays}c and Fig.~\ref{fig:halpha}) 
of high surface brightness embedded in a more 
irregular complex of extended emission. The part of F1 near the galaxy nucleus is extended.
At the western part of F1, H$\alpha$ blobs of higher surface brightness might
indicate the presence of H{\sc ii} regions. In addition, two high surface brightness filaments 
(F3 and F4) are located southwest of the nucleus.
The high [N{\sc ii}]/H$\alpha$ ratio observed in F3
(Fig.~6 of Kenney et al. 1995) indicates that the diffuse ionized gas located in F1--F4,
which have high surface brightnesses, is shock excited.
Chemin et al. (2005) detected a new H$\alpha$ filament F0 (Fig.~\ref{fig:halpha}) north of the extraplanar 
emission observed by Kenney et al. (1995). They describe this filament as a string of H$\alpha$ knots
(probably H{\sc ii} regions) embedded in faint diffuse emission. 
The very deep narrow band image of Kenney et al. (2008) shows these H{\sc ii} regions and
a very small amount of extended emission, in clear contrast to filaments F1--4.
Filament F0 shows velocities close to the systemic
velocity of NGC~4438. Thus, in general, the extraplanar warm ionized gas observed in H$\alpha$ consists 
mainly of a diffuse phase with embedded dense H{\sc ii} regions. The extended high-surface brightness gas is 
probably shock-excited. 
Future [N{\sc ii}]/H$\alpha$ line ratios of this region are needed to confirm this conclusion.
Machacek et al. (2004) showed that the X-ray emission closely follows the high surface 
brightness H$\alpha$ emission distribution. There is no X-ray emission associated with the H$\alpha$ filament F0,
consistent with ionization by massive stars.

The multiwavelength comparison shows the following:
\begin{itemize}
\item
Only the eastern half of the 6~cm extraplanar region is detected in H$\alpha$.
This supports our conclusion that the 6~cm radio continuum emission is mostly non-thermal.
\item
There is no radio counterpart of the southernmost H$\alpha$ filament. 
\item
High surface brightness H$\alpha$ is emitted from a region where CO and radio continuum emission overlap 
west of the galaxy nucleus. 
\item
The extraplanar CO emission extends $\sim 30''$ ($2.5$~kpc) farther to the north
than the high surface brightness H$\alpha$ and the 6~cm radio continuum emission (F0). 
\item
The northern extraplanar H{\sc ii} filament F0 coincides with the H{\sc i} emission peak.
There, the H$\alpha$ velocity is consistent with the H{\sc i} velocity (Chemin et al. 2005). 
\item
The extraplanar H$\alpha$ filament F1 does not have a FUV counterpart and Fig.~\ref{fig:halpha}
shows that [N{\sc ii}] probably dominates the narrow-band emission.
\end{itemize}
We conclude that the northern part is different from the rest of the extraplanar region.
In the former, high surface brightness 
H{\sc i} and faint H$\alpha$ emission mainly in the form of H{\sc ii} regions is found, 
whereas in the latter, high surface density H$\alpha$, 
X-ray, and 6~cm radio continuum emission is present.
In addition, the H$\alpha$ filament F1 has no FUV counterpart, i.e. it is devoid of stars younger than $100$~Myr.
We attribute the extended low surface brightness H$\alpha$ emission of F0 to star formation whereas
the extended high surface brightness H$\alpha$ emission is most probably due to shocks.

The absence of 6~cm radio continuum, extended high surface brightness H$\alpha$, and
X-ray emission in the
northern part of the extraplanar region (F0) is probably because the diffuse warm ionized and magnetic 
fields/cosmic ray electrons are stripped more efficiently than the warm and cold dense neutral gas 
traced by CO and H{\sc i} emission.
The H{\sc i} and faint H$\alpha$ emission in F0 might be interpreted as the outer rim of the molecular clouds
which are exposed to the radiation field of the galaxy (UV) and the intracluster medium (X-rays).
This leads to the formation of ionized rims and photodissociation regions where atomic hydrogen
can be found. In this scenario, the ionized and neutral gas is physically linked to dense gas and
should show the same kinematics as the molecular gas. 
This is true for the H$\alpha$ and H{\sc i} emitting gas (Chemin et al. 2005)
and for the H$\alpha$ and CO emitting gas (Sect.~\ref{sec:kinematics}).
On the other hand, the H$\alpha$ filament F1 whose eastern part close to the galaxy
nucleus is devoid of H{\sc ii} regions might not be linked to dense gas. 
If the diffuse warm ionized gas is of low density ($< 10$~cm$^{-3}$ instead of $\ga 100$~cm$^{-3}$
for the neutral gas) we expect it to be stripped more efficiently by the intracluster medium ram pressure.
In this case the H$\alpha$ kinematics should be different from the CO kinematics.
This will be investigated in the next section.

\section{Kinematic evidence of more efficient stripping of ionized gas \label{sec:kinematics}}

\begin{figure*}
        \resizebox{\hsize}{!}{\includegraphics{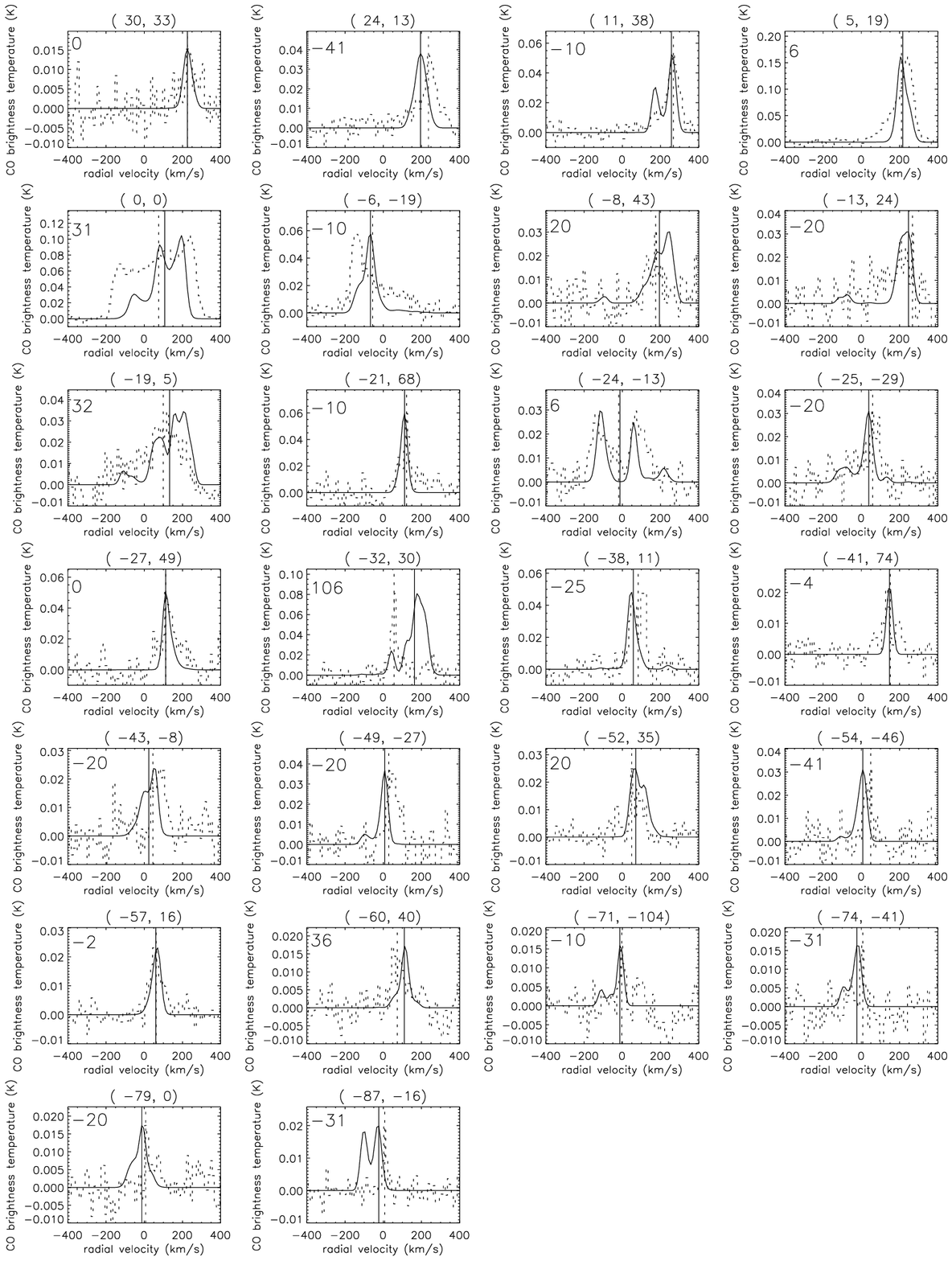}}
        \caption{Solid line: H$\alpha$ spectra generated from Chemin et al. (2005), dashed line: CO(1--0) spectra
	  from Vollmer et al. (2005). Vertical lines: maxima or barycenters of the
	  line profiles. The peak of the H$\alpha$ line is normalized to the peak of the CO(1--0) line.
	  The positions with respect to the galaxy center are plotted on top of each panel.
	  The velocity differences are indicated in the upper left corner of each panel.
        } \label{fig:spectra}
\end{figure*}

To investigate our hypothesis of more efficient stripping of the diffuse ionized phase of the
ISM, we compare CO(1-0) and H$\alpha$ kinematics.
We took the CO(1-0) spectra with a resolution of $21''$ from Vollmer et al. (2005).
Since the H$\alpha$ Fabry-Perot data cube of Chemin et al. (2005) is contaminated
by night sky line emission, we reconstructed for each CO(1--0) position an H$\alpha$ spectrum
using the first and second moment maps of the Fabry-Perot data
and assuming a constant linewidth of $30$~km\,s$^{-1}$. 
The overlays of all CO(1--0) with the regenerated H$\alpha$ spectra are shown in 
Fig.~\ref{fig:spectra}.
In general, there is good agreement between the CO(1--0) and H$\alpha$ line profiles.
In particular,  the double line profile which is characteristic of ram pressure stripping and which is
detected in position $(-24,-13)$ is also present in H$\alpha$ (Fig.~\ref{fig:spectra}). 

\begin{figure}
        \resizebox{\hsize}{!}{\includegraphics{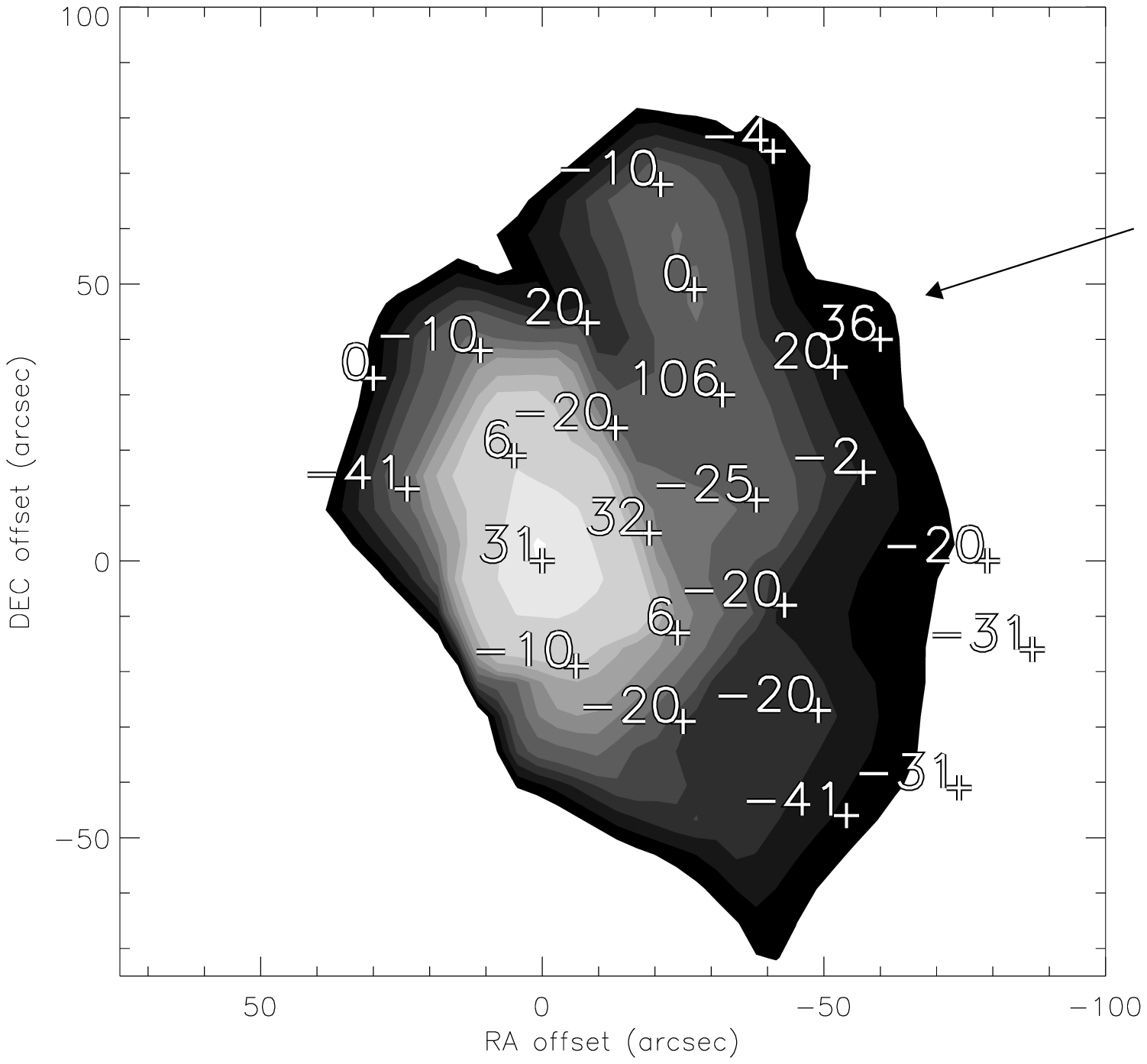}}
        \caption{Difference between the H$\alpha$ and CO(1--0) velocities on the CO(1--0)
	  emission map from Vollmer et al. (2005). The arrow indicates a region of
	  unusually large velocity differences. The mean radial velocity of the Virgo cluster
	  is $1050$~km\,s$^{-1}$.
        } \label{fig:mom0}
\end{figure}

\begin{figure}
        \resizebox{\hsize}{!}{\includegraphics{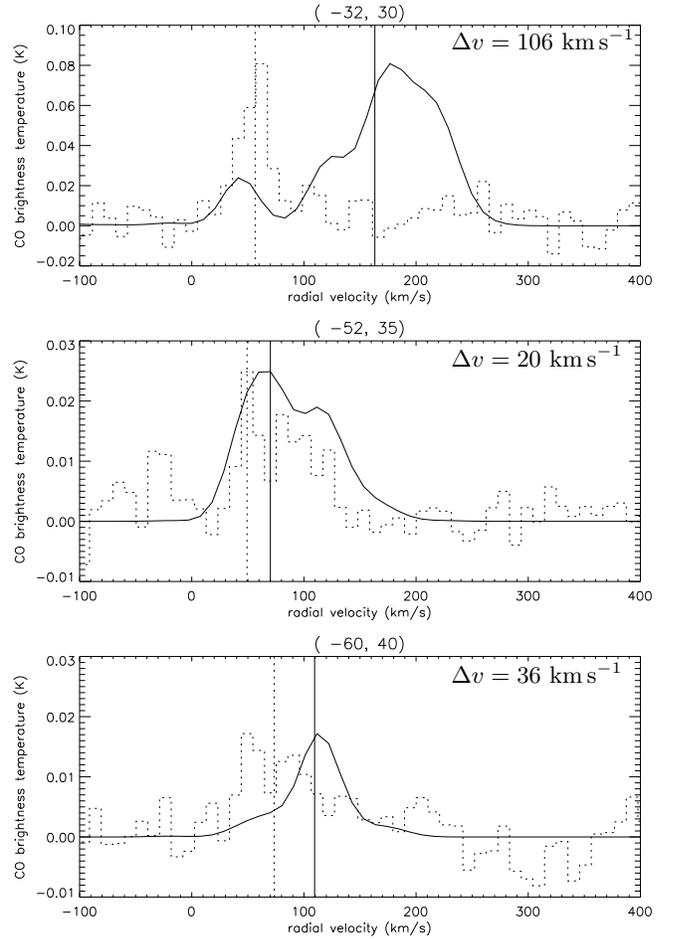}}
	\put(-80,340){$\Delta v= 106$~km\,s$^{-1}$}
	\put(-80,220){$\Delta v= 20$~km\,s$^{-1}$}
	\put(-80,100){$\Delta v= 36$~km\,s$^{-1}$}
        \caption{Spectra corresponding to the locations indicated by an arrow in Fig.~\ref{fig:mom0}.
	  Solid line: H$\alpha$ spectra from Chemin et al. (2005), dashed line: CO(1--0) spectra
	  from Vollmer et al. (2005). Vertical lines: maxima or barycenters of the
	  line profiles. The peak of the H$\alpha$ line is normalized to the peak of the CO(1--0) line.
	  The positions with respect to the galaxy center are plotted on top of each panel.
        } \label{fig:kinematics}
\end{figure}

To determine a unique velocity difference between the H$\alpha$ and CO(1--0) spectra we
determined the differences between (i) the maxima and (ii) the barycenters of the lines.
We then conservatively took the smallest velocity difference.
In Fig.~\ref{fig:mom0} we show these differences on the CO(1--0) emission map.
In the galactic disk the relatively high velocity differences of $-40$ and $+30$~km\,s$^{-1}$
are not reliable due to the complex line profiles and effects of H$\alpha$ absorption
(see Fig.~\ref{fig:spectra}).
In the other regions, the absolute of the velocity difference do not exceed $31$~km\,s$^{-1}$
except in a region northwest from the galactic nucleus which corresponds to the
H$\alpha$ filament F1 indicated with an arrow on Fig.~\ref{fig:mom0}.
These three spectra are shown in Fig.~\ref{fig:kinematics}.
Whereas the velocity difference is small at $(-52,35)$ it is somewhat higher than average ($36$~km\,s$^{-1}$) 
at $(-60,40)$ and unusually high ($106$~km\,s$^{-1}$) at $(-32,30)$. At the latter position, there is 
also a weak H$\alpha$ line associated with the CO(1--0) line. The high H$\alpha$ velocity
at $(-32,30)$ can be seen as an extraplanar violet region in the H$\alpha$ velocity field 
(Fig.~2 of Chemin et al. 2005).
The velocity difference is positive, i.e. towards the velocity of the Virgo cluster ($1050$~km\,s$^{-1}$).
Therefore, this provides more evidence of our suggestion that the diffuse ionized gas in the
eastern part of F1 is stripped more efficiently than the warm and cold dense neutral gas.
We interpret the faint H$\alpha$ line associated with the CO line as due to star formation.
It represents a natural continuation of the faint H$\alpha$ emission detected in F0 and
is associated with the H{\sc i} emission of lower surface brightness.
The large velocity difference occurs in the eastern part of F1. In the deep narrow band image of Kenney et al. (2008)
its elongation is roughly east--west, the same direction as the northern very faint H$\alpha$ filament
at a declination of $13^{\circ}06'30''$. This common direction corresponds exactly to the projected 
ram pressure wind angle in the model of Vollmer et al. (2005) and thus represents additional
evidence of ongoing strong ram pressure acting on NGC~4438's ISM.

\section{A unified picture of multiphase ram pressure stripping \label{sec:unified}}

Whereas the origin of NGC~4438's tidal interaction remains unclear (NGC~4435 or M~86),
the interaction timescale is $\sim 100$~Myr for both scenarios. Despite this uncertainty, the
CO and H$\alpha$ distribution and kinematics are consistent with and naturally explained by
strong ongoing ram pressure stripping of NGC~4438's ISM.

Different ram pressure stripping efficiencies of different ISM phases can be understood
in terms of gas column densities. The acceleration of a gas clump of surface density $\Sigma$ 
due to ram pressure $p_{\rm ram}$ is $a = p_{\rm ram}/\Sigma$.
The surface density $\Sigma$ of a spherical cloud of mass $M_{\rm cl}$ and radius $R_{\rm cl}$ is 
$\Sigma=M_{\rm cl}/(\pi R_{cl}^2)=4/3\,\rho_{\rm cl} R_{\rm cl}$. 
Thus, when a gas clump is heated and expands at constant mass, the surface density drops 
and the acceleration due to ram pressure increases significantly.
We think that this is the basic mechanism for the observed spatial and velocity offset
between the high surface brightness H$\alpha$ and CO(1--0) emission.

Other observations of Virgo spiral galaxies have already been interpreted in this sense:
\begin{itemize}
\item
In the northern tidal arm of NGC~4438, which is devoid of atomic gas, Vollmer et al. (2005)
detected molecular gas at the velocities of the stars in the tail. These authors
interpreted their finding as a decoupling of the dense molecular clouds from the
ram pressure wind. 
\item
In NGC~4402 Crowl et al. (2005) observed numerous H{\sc ii} complexes along the southern edge 
of the gas disk, possibly indicating star formation triggered by the intracluster 
medium (ICM) pressure.
To the south of the main ridge of interstellar material, where the galaxy is relatively clean of gas 
and dust, 1~kpc long linear dust filaments were discovered with a position angle that matches the 
extraplanar radio continuum tail. One of the observed dust filaments has an H{\sc ii} region at its head. 
These dust filaments were interpreted as large, dense clouds that were initially left behind as the 
low-density interstellar medium was stripped but were then ablated by the ICM wind.
\item
The peculiar velocity of the compact extraplanar H{\sc ii} region to the north of NGC~4388
(Gerhard et al. 2002) can be explained by a scenario where a stripped gas cloud decoupled
from the wind, forms stars, and falls back onto the galactic disk (see Vollmer \& Huchtmeier 2003).
\end{itemize}
All these observations indicate that the stripping efficiency depends on the surface density
of the stripped gas clumps.

As for the ram pressure stripping of the cosmic ray gas with its associated magnetic field,
the pressure based on the assumption of minimum energy is $p=2 \times 10^{-12}$~dyn\,cm$^{-2}$
for a magnetic field strength of $10$~$\mu$G.
This is a factor of $30$ lower than the current ram pressure derived from the dynamical
model of Vollmer et al. (2005). Therefore, the cosmic ray gas and its associated magnetic field
are easily stripped together with the diffuse warm/hot ionized interstellar medium.
The spatial coincidence between the H$\alpha$/X-ray and radio continuum emission in the eastern half
of the extraplanar region tells us that the stripping efficiency for the cosmic ray gas
and its associated magnetic field
is not significantly different from that of the diffuse warm/hot ionized gas.
This is consistent with the picture that the galactic magnetic fields are coupled to the diffuse
ionized medium (Beck et al. 2005) and not to the dense neutral instellar medium.
The observed velocity offset between the two gas phases cannot be explained in 
a scenario where the magnetic fields are anchored in the dense neutral gas and pervade the adjacent
ionized gas, because the less massive ionized gas would have to follow the dense neutral gas.

Murphy et al. (2008a, 2008b) compared Spitzer $24$~$\mu$m emission with 20~cm
radio continuum maps. The radio-FIR correlation is used to predict the radio
emission from the Spitzer $24$~$\mu$m emission. They found radio deficient regions
at the outer edges of the galactic disks where ram pressure is pushing the
interstellar medium. Since the $24$~$\mu$m dust emission is associated with molecular gas,
the northern part of the extraplanar CO(1--0) emission can be qualified as radio deficient.
We therefore suggest that the radio-deficient regions are due to the more efficient
ram pressure stripping of diffuse ionized gas (thermal and cosmic ray gas with its associated magnetic field) 
with respect to the dense molecular gas.

\begin{acknowledgements}
This research has made use of the GOLD Mine Database.
This work was supported by the Polish-French (ASTRO-LEA-PF) cooperation program
and by the Polish Ministry of Sciences and Higher Education grant 2693/H03/2006/31.
\end{acknowledgements}



\end{document}